\documentclass[
aps,
prx,
reprint,
superscriptaddress,
longbibliography,
]{revtex4-2}
\usepackage{amsmath,amsfonts,graphicx,color,bbm,tikz,bm,setspace}
\usepackage{graphicx,booktabs}
\usepackage{hyperref}
\usepackage{float}
\usepackage{bbold}
\usepackage{subfig}
\usepackage{multirow}
\usepackage[utf8]{inputenc}
\usepackage[abs]{overpic}
\usepackage{xcolor,varwidth}
\usepackage{tikz, pgfplots}
\usepackage[dvipsnames]{xcolor}
\usetikzlibrary{positioning}
\usetikzlibrary{decorations.pathreplacing}
\usepackage{physics}
\usetikzlibrary{angles,quotes,calc}
\usepackage{amsmath,amsfonts,graphicx,color,bbm,amssymb,amsthm}
\usetikzlibrary{decorations.pathreplacing,arrows.meta,positioning,shapes.geometric,fadings}

\begin{document}

\title{When Is Kramers-Wannier Duality Invertible? }

\author{Akash Sinha}
\email{akash26121999@gmail.com}
\affiliation{\it Department of Physics, School of Basic Sciences,\\ Indian Institute of Technology, Bhubaneswar, 752050, India}

\author{Pramod Padmanabhan}%
\email{pramod23phys@gmail.com,}
\affiliation{\it Department of Physics, School of Basic Sciences,\\ Indian Institute of Technology, Bhubaneswar, 752050, India}


\author{Vladimir Korepin}
\email{vladimir.korepin@stonybrook.edu}
\affiliation{C. N. Yang Institute for Theoretical Physics, \\ Stony Brook University, New York 11794, USA}

\begin{abstract}
     {\it Kramers–Wannier} duality of the quantum Ising chain is naturally realized as a noninvertible transformation on a finite periodic chain. Upon incorporating appropriate symmetry-twisted sectors, however, the duality can be promoted to an invertible unitary transformation. In this Letter, we classify when the Kramers-Wannier duality admits an invertible realization on a finite Hilbert space. We show that this is determined by the representation of the Ising bond algebra and establish a necessary and sufficient condition in terms of the traces of its two central elements. This representation-theoretic criterion unifies the conventional noninvertible realization with invertible constructions involving symmetry-twisted sectors and applies equally to other models with the same underlying bond algebra. As an explicit example, we find a hidden order-disorder duality in a non-trivial realization of the Ising bond algebra that admits an invertible Kramers–Wannier duality.
\end{abstract}

\maketitle

\section{Introduction.}
Dualities \cite{Savit1981,PRXQuantum.4.020357,PRXQuantum.5.010338} relate distinct descriptions of quantum many-body systems, connecting phases, spectra and correlation functions. In particular, they can reveal critical points and phase boundaries without requiring a complete solution of the model. A canonical example is the \textit{Kramers-Wannier} (KW) duality \cite{KW1941-1,KW1941-2} of the $2$-D classical Ising model, which in the one-dimensional quantum Ising chain \cite{lieb1961two,katsura1962statistical,niemeijer1967some,pfeuty1970one,elliott1970ising} manifests as a strong-weak duality, exchanging the symmetry-broken ferromagnetic and symmetric paramagnetic phases. A natural operator-level formulation of such dualities is provided by the \textit{bond algebra} \cite{PhysRevB.79.214440,PhysRevLett.104.020402, cobanera2011bond}, which focuses on the algebra generated by the local terms of a Hamiltonian. Rather than specifying the transformation of microscopic variables, a duality can be formulated as a map between bond algebras that preserves their defining algebraic relations. This perspective is particularly useful for the Ising model, where the bond algebra \cite{minami2016solvable,ogura2020geometric,sinha2026hidden} is generated by the hermitian densities ${h_j}=h_j^\dagger$, which anticommute with the nearest neighbors and commute with the rest as
\begin{eqnarray}\label{eq:IsingBondAlgebra}
\{h_j,h_{j+1}\}=0,\quad [h_j,h_k]=0,~|j-k|\neq1,\quad h_j^2=1,
\end{eqnarray}
with $j,k=1,\ldots,2N$ and the identification $h_{2N+1}\cong h_1$. One recovers the periodic quantum Ising chain by choosing $h_{2j-1}=Z_j$ and $h_{2j}=X_jX_{j+1}$, with $X_{N+1}\equiv X_1$. The KW duality exchanges the strong and weak-coupling regimes of the Ising Hamiltonian $H_{\rm I}(g)=-g\sum_j Z_j-\sum_j X_jX_{j+1}$ through the mapping $H_{\rm I}(g)\to H_{\rm I}(1/g)$. On a finite periodic chain, however, this transformation is not invertible on the full Hilbert space \cite{grimm1993spin,aasen2016topological,aasen2020topological,Zhang_2025,Sinha:2025wqf,Seiberg2024_NonInvertibleLSM,seiberg2024majorana,Seiberg2024_NonInvertibleLSM,Zhang_2025} and becomes an exact reversible map only after restriction to an appropriate symmetry sector. The same bond algebra can instead be realized on $(N+1)$ spins by modifying the boundary density to $h_{2N}=X_NZ_0X_1$. The auxiliary spin gives rise to the conserved operator $Z_0$, with its $\pm 1$ eigenvalue sectors realizing the periodic and antiperiodic $N$-site Ising chains, respectively. In this way, the enlarged representation incorporates both the periodic and symmetry-twisted sectors within a single Hilbert space. Such an enlargement is known to promote the KW duality to an invertible transformation acting on the full Hilbert space\cite{PRXQuantum.4.020357,PRXQuantum.5.010338,PhysRevB.108.214429,seiberg2024majorana,Seiberg2024_NonInvertibleLSM,PhysRevLett.134.130403}.

Therefore, although the underlying Ising bond algebra remains the same, different representations of it can lead to both invertible and noninvertible KW dualities. In this work, we consider the Hamiltonian
\begin{eqnarray}\label{eq:GenIsingBonHam}
H(g)=-g\sum_{j=1}^N h_{2j-1}-\sum_{j=1}^N h_{2j},
\end{eqnarray}
in a general finite-dimensional representation of the algebra generated by $\{h_j\}$, where the generators satisfy \eqref{eq:IsingBondAlgebra}. The KW duality exchanges the strong and weak-coupling regimes of the above Hamiltonian. At the level of the Hamiltonian densities, the above mapping can be realized as
\begin{eqnarray}\label{eq:invertibleKW}
{\cal D}h_{j}=h_{j+1}{\cal D},
\end{eqnarray}
so that the duality ${\cal D}$ shifts the local densities while preserving their defining algebraic relations. This transforms the total Hamiltonian \eqref{eq:GenIsingBonHam} as ${\cal D}{H(g)}=gH\left(1/g\right){\cal D}$. Crucially, we require the duality operator ${\cal D}$ to act within a single Hilbert space ${\cal H}$, carrying a definite representation of the bond algebra.
Several invertible realizations of KW duality have been established previously in specific settings of the quantum Ising chain, including formulations with nonlocal boundary terms \cite{cobanera2011bond,seiberg2024majorana,Sinha:2025wqf}; constrained Hilbert spaces without a local tensor product structure \cite{jones2021one}; enlarged Hilbert spaces incorporating symmetry-twisted
sectors \cite{PhysRevB.108.214429,seiberg2024majorana,Seiberg2024_NonInvertibleLSM}. {More broadly, recent {\it matrix product operator} (MPO) and categorical constructions \cite{PRXQuantum.4.020357,PRXQuantum.5.010338} show how one-dimensional dualities can be promoted to unitary linear-depth quantum circuits on enlarged Hilbert spaces by introducing ancillary degrees of freedom \cite{PhysRevLett.134.130403}.} In this letter, we adopt a general representation-theoretic perspective and ask when KW duality is invertible in an arbitrary finite-dimensional representation of the Ising bond algebra, without assuming additional structure such as particular boundary conditions or the presence of ancillary degrees of freedom. We provide a complete answer to this question. We show that invertibility is controlled by the representation of the bond algebra and derive a necessary and sufficient condition in terms of the two operators
\begin{eqnarray}
    P_{\rm o}=\prod_{j=1}^Nh_{2j-1},~~ P_{\rm e}=\prod_{j=1}^Nh_{2j},\quad [P_{\rm o,\rm e},h_j]=0.
\end{eqnarray}
Both of them are hermitian $P_{\rm o,\rm e}^\dagger=P_{\rm o,\rm e}$ and square to identity $P_{\rm o,\rm e}^2=1$. Since they commute with each other as well as with all the Hamiltonian densities, they belong to the center of the bond algebra. As a result the Hilbert space decomposes into the invariant sectors
\begin{eqnarray}\label{eq:invSec}
    {\cal H}={\cal H}_{++}\oplus{\cal H}_{--}\oplus{\cal H}_{+-}\oplus{\cal H}_{-+},
\end{eqnarray}
labeled by the eigenvalues of $P_{\rm o,\rm e}$, respectively. The duality exchanges these two central elements as ${\cal D}P_{\rm o,\rm e}{\cal D}^{-1}=P_{\rm e,\rm o}$ and hence maps ${\cal D}:{\cal H}_{\pm\pm}\to{\cal H}_{\pm\pm},\,{\cal D}:{\cal H}_{\pm\mp}\to{\cal H}_{\mp\pm}$. Therefore, a \textit{necessary} condition for the existence of an invertible duality is 
\begin{eqnarray}\label{eq:tracecond}
    {\rm Tr}~P_{\rm o}={\rm Tr}~P_{\rm e}.
\end{eqnarray}
In particular, this ensures that the exchanged sectors ${\cal H}_{\pm\mp}$ have equal dimensions, as required for an invertible mapping. In this letter, we prove that the condition \eqref{eq:tracecond} is also \textit{sufficient} for an invertible duality. Our result places a range of finite-size realizations of KW duality within a unified representation-theoretic framework. What appear as distinct invertible and noninvertible realizations are thereby understood as different representations of the same Ising bond algebra, with their invertibility determined by the single criterion $\mathrm{Tr}\,P_{\rm o}=\mathrm{Tr}\,P_{\rm e}$. Interestingly, imposing additional constraints on the representation leads to stronger consequences for the structure of the duality, which we summarize in Table \ref{tab:KWhierarchy}. Furthermore, when the trace criterion is violated, we show how to enlarge the Hilbert space by a finite representation extension that removes the obstruction and restores an invertible KW duality.


\section{Invertible KW duality.}
\label{sec:noninvdual}
Let us first address the question of when the duality can be implemented by an invertible element in the bond algebra itself. As we know, the duality exchanges the central elements as ${\cal D}P_{\rm o,\rm e}{\cal D}^{-1}=P_{\rm e,\rm o}$. However, if an invertible duality operator ${\cal D}$ belongs to the bond algebra, it must commute with every element of its center. Hence, ${\cal D}P_{\rm o,e}=P_{\rm o,e}{\cal D}$ which is compatible with an invertible duality only when $P_{\rm o}=P_{\rm e}$. Therefore, the equality $P_{\rm o}=P_{\rm e}$ is a necessary condition for realizing the duality as an invertible operator within the bond algebra. When $P_{\rm o}\neq P_{\rm e}$, their exchange under duality constitutes an algebraic obstruction to such an invertible realization. Nonetheless, there exists an operator in the bond algebra which always implements the duality transformation $U_+:h_j\to h_{j+1}$ in the bulk
\begin{eqnarray}\label{eq:unitarypreKW}
    U_+=\prod_{j=1}^{2N-1}\frac{1+{\rm i}h_j}{\sqrt{2}},\qquad U_+^\dagger U_+=1.
\end{eqnarray}
The unitary $U_+$ may be interpreted as a {\it sequential quantum circuit} \cite{PhysRevB.109.075116} whose depth scales linearly with the system size.
Crucially, $U_+$ acts on the local Hamiltonian densities as
\begin{eqnarray}
    &U_+h_jU_+^{-1}=h_{j+1},\qquad j\neq 2N-1,2N,\nonumber\\
    &U_+h_{2N-1}U_+^{-1}=\hat{\cal P}h_{2N},\quad U_+h_{2N}U_+^{-1}=\hat{\cal P}h_{1},
\end{eqnarray}
where $\hat{\cal P}$ is a global parity operator, defined by
\begin{eqnarray}
    \hat{\cal P}=P_{\rm o}P_{\rm e}=\prod_{j=1}^N h_{2j-1}\prod_{j=1}^N h_{2j}={\cal P}^\dagger,\quad {\cal P}^2=1.
\end{eqnarray}
In the thermodynamic limit $N\to\infty$, the unitary $U_+$ rightly describes (at least formally) the duality operator that shifts each local Hamiltonian density by one index $U_+:h_j\to h_{j+1}$. However, unless $\hat{\cal P}$ is identity on the full Hilbert space, this mapping is not exact on a finite chain as the local boundary Hamiltonian densities are mapped to non-local operators. 
Nonetheless, when restricted to the global parity-even sector $(\hat{\cal P}=1)$, the unitary $U_+$ coincides with the required ${\cal D}$, letting us introduce
\begin{eqnarray}\label{eq:noninvKWD}
    &{\cal D}_{+}:=\hat{\cal P}_+U_+\hat{\cal P}_+,\quad \hat{\cal P}_\pm:=(1\pm\hat{\cal P})/2,\\
    &{\cal D}_{+}h_j=h_{j+1}{\cal D}_{+},~~\hat{\cal P}_+{\cal D}_+\hat{\cal P}_+={\cal D}_+,~~ {\cal D}_+^\dagger{\cal D}_+=\hat{\cal P}_+.\nonumber
\end{eqnarray}
Here $\hat{\cal P}_\pm$ denote projectors to the even and odd-parity sectors, respectively. The operator ${\cal D}_+$ annihilates all the parity odd $(\hat{\cal P}=-1)$ states and acts as the unitary KW duality when restricted to the parity even sector. Furthermore, being noninvertible in general, ${\cal D}_{+}$ cannot act by conjugation. 
Rather, its action is defined through \eqref{eq:noninvKWD}.
We emphasize that the noninvertibility of the above duality symmetry is directly tied to the assumption that $\hat{\cal P}$ does not act as the identity on the full Hilbert space. Conversely, when $\hat{\cal P}$ becomes identity, i.e. $P_{\rm o}=P_{\rm e}$, we recover $U_+$ from ${\cal D}_{+}$ as the corresponding invertible duality operator. This verifies the earlier observation that an invertible duality can be found in the bond algebra if $P_{\rm o}=P_{\rm e}$. In general, however, $\hat{\cal P}$ is not identity, and ${\cal D}_{+}$ is therefore noninvertible by construction. 


To extend the duality in the full Hilbert space, we now seek a parity-preserving operator ${\cal D}_-$, satisfying
\begin{eqnarray}\label{eq:noninvKWDP-}
    {\cal D}_{-}h_j=h_{j+1}{\cal D}_{-},~~\hat{\cal P}_-{\cal D}_-\hat{\cal P}_-={\cal D}_-,~~{\cal D}_-^\dagger{\cal D}_-=\hat{\cal P}_-.
\end{eqnarray}
In other words, ${\cal D}_-$ annihilates all the parity-even states and implements the unitary duality transformation in the odd-parity sector. One can then glue ${\cal D}_\pm$ sector-wise to construct
\begin{eqnarray}\label{eq:KWop}
    &{\cal D}=\hat{\cal P}_+{\cal D}_+\hat{\cal P}_++\hat{\cal P}_-{\cal D}_-\hat{\cal P}_-={\cal D}_++{\cal D}_-,\nonumber\\
    &{\cal D}^\dagger{\cal D}=1,\quad[{\cal D},\hat{\cal P}]=0,\quad{\cal D}h_j{\cal D}^\dagger=h_{j+1},
\end{eqnarray}
thus implementing the unitary KW duality over the full Hilbert space. Notably, ${\cal D}_\pm$ can be multiplied by arbitrary $U(1)$ phase factors without harming the duality. We now show that Eq.~\eqref{eq:tracecond} is \textit{sufficient} for the existence of ${\cal D}_-$. To this end, consider the hermitian operators
\begin{eqnarray}\label{eq:auxferm}
    f_j={\rm i}^{j-1}\prod_{k=1}^j h_k,\quad f_j^\dagger=f_j,\quad j=1,\cdots,2N-1.
\end{eqnarray}
This allows the Hamiltonian densities to be expressed as $h_1=f_1,\,h_j=-{\rm i}f_{j-1}f_{j},\,j\neq 1,2N$. Since $\{f_j\}$ are constructed entirely out of the Hamiltonian densities, all of them commute with the operators $P_{\rm o,\rm e}$. In particular, one finds $P_{\rm o}=(-{\rm i})^{N-1}f_1\cdots f_{2N-1}$. They are odd in number and satisfy
\begin{eqnarray}\label{eq:auxfermrel}
    \{f_j,f_k\}=2\delta_{j,k},\qquad [f_j,P_{\rm o,\rm e}]=0,
\end{eqnarray}
with $j,k=1,\cdots,2N-1$. Since \eqref{eq:auxfermrel} directly follows from the relations \eqref{eq:IsingBondAlgebra}, the physical Hilbert space ${\cal H}$ carrying a representation of the bond algebra, also furnishes a representation of the Clifford algebra generated by $\{f_j\}$. Recall that, the two central elements $P_{\rm o,\rm e}$ help decompose the total Hilbert space into the invariant sectors $\{{\cal H}_{\alpha\alpha'}|\alpha,\alpha'=\pm\}$, labeled by the eigenvalues of $P_{\rm o,\rm e}$. Imposing ${\rm Tr}\,P_{\rm o}={\rm Tr}\,P_{\rm e}$ \eqref{eq:tracecond} now ensures that the sectors $\mathcal H_{+-}$ and $\mathcal H_{-+}$ have equal dimension. As we rigorously establish in App. \ref{App:OddMajorana}, this allows the construction of a hermitian operator $\eta_-=\eta_-^\dagger$ that interchanges ${\cal H}_{+-}$ and ${\cal H}_{-+}$, while anticommutes with the Clifford generators and the central operators as
\begin{eqnarray}\label{eq:chirela}
    &\{\eta_-,f_j\}=0=\{\eta_-,P_{\rm o,\rm e}\},\nonumber\\
    &\hat{\cal P}_-\eta_-\hat{\cal P}_-=\eta_-,\quad \eta_-^\dagger\eta_-=\eta_-\eta_-^\dagger=\hat{\cal P}_-.
\end{eqnarray}
Consequently, $\eta_-$ anticommutes with $h_{1},h_{2N}$ and commutes with the rest: $\{\eta_-,h_{1,2N}\}=0,~[\eta_-,h_j]=0,\,j\neq1,2N$. We can then define
\begin{eqnarray}
    {\cal D}_-=\hat{\cal P}_-\left(\eta_- U_+\right)\hat{\cal P}_-,
\end{eqnarray}
which satisfies all the requirements in \eqref{eq:noninvKWDP-}, allowing us to define the unitary duality \eqref{eq:KWop} and hence enabling an invertible KW duality within the full Hilbert space. This establishes the sufficiency of the condition \eqref{eq:tracecond} for the existence of an invertible KW duality.

Interestingly, if we further impose ${\rm Tr}\,P_{\rm o}=0={\rm Tr}\,P_{\rm e}$, the sectors ${\cal H}_{\pm\pm}$ also have equal dimension \eqref{App:OddMajorana} and $\eta_-$ can be promoted to a unitary $\eta$ with $\eta^\dagger\eta=1$, which again satisfies a relation analogous to \eqref{eq:chirela} . The operator $\eta_-$ is then recovered by projection  onto the odd-parity sector $\eta_-=\hat{\cal P}_-\eta\hat{\cal P}_-$. In such a scenario, we may define the fermionic operators $\gamma_1=\eta,\gamma_j={\rm i}\eta f_{j-1},~j=2,\cdots,2N$, which satisfy $\{\gamma_j,\gamma_k\}=2\delta_{j,k}$. The Hamiltonian densities now can be expressed as the fermionic bilinears $h_j=-{\rm i}\gamma_j\gamma_{j+1},j\neq 2N,~h_{2N}={\rm i}\hat{\cal P}\gamma_{2N}\gamma_1$. An important consequence is that, at the self-dual point $g=1$, the Hamiltonian \eqref{eq:GenIsingBonHam} becomes Yang-Baxter integrable \cite{Sinha:2025wqf,sinha2026noninvertible}. In particular, the unitary operator $U_+$ in \eqref{eq:unitarypreKW} admits a construction via the \textit{quantum inverse-scattering method}\cite{Korepin1993QuantumIS,faddeev1996algebraic}, in terms of a well-defined $R$-operator. 


\begin{table}[h!]
\centering
\caption{\label{tab:KWhierarchy}
Representation-theoretic hierarchy of finite-size Kramers-Wannier duality.
}
\begin{ruledtabular}
\begin{tabular}{p{0.32\columnwidth} p{0.65\columnwidth}}
{\bf Representation condition}
& \hspace{1.8cm}{\bf KW duality}
\\ \hline
$\Tr P_{\rm o}=\Tr P_{\rm e}$
&
Necessary and sufficient for invertible realization on the full Hilbert space $\mathcal{H}={\cal H}_{++}\oplus{\cal H}_{--}\oplus{\cal H}_{+-}\oplus{\cal H}_{-+}$, with ${\rm dim}{\cal H}_{+-}={\rm dim}{\cal H}_{-+}$. Admits an intertwiner $\eta_-:{\cal H}_{\pm\mp}\to{\cal H}_{\mp\pm}$, satisfying $\eta_-^\dagger\eta_-=\hat{\cal P}_-$.
\\[4pt]

$\Tr P_{\rm o}=\Tr P_{\rm e}=0$
&
Invertible on $\mathcal{H}$ with additionally ${\rm dim}{\cal H}_{++}={\rm dim}{\cal H}_{--}$. Noninvertible $\eta_-$ can be extended to a unitary $\eta$ with $\eta^\dagger\eta=1$, yielding a free-fermion realization.
\\[4pt]

$P_{\rm o}=P_{\rm e}$
&
Necessary and sufficient for the duality operator as an invertible element of the bond algebra itself.
\\[4pt]

\end{tabular}
\end{ruledtabular}
\end{table}

The trace criterion also provides a constructive way to remove the obstruction to an invertible duality. 
Suppose that a representation $\{h_j\}$ satisfies ${\rm Tr}\,P_{\rm o}-{\rm Tr}\,P_{\rm e}\neq0$. We may enlarge it by adjoining $\ell=|{\rm Tr}\,P_{\rm o}-{\rm Tr}\,P_{\rm e}|/2^N$ copies of an $N$-site Ising representation,
\begin{eqnarray}
    &\tilde{h}_{2j-1}=h_{2j-1}\oplus Z_j^{\oplus\ell},
    \quad j=1,\ldots,N,\\
    &\tilde{h}_{2j}=h_{2j}\oplus (X_jX_{j+1})^{\oplus\ell},
    \quad j=1,\ldots,N-1,\nonumber\\
    &\tilde{h}_{2N}=h_{2N}\oplus
    (\varepsilon X_NX_1)^{\oplus\ell},\quad \varepsilon={\rm sgn}({\rm Tr}\,P_{\rm o}-{\rm Tr}\,P_{\rm e}).\nonumber
\end{eqnarray}
Each additional block realizes the same Ising bond algebra. Moreover, we have ${\rm Tr}\,Z_1\cdots Z_N=0$, whereas $(X_1X_2)(X_2X_3)\cdots
(X_{N-1}X_N)(\varepsilon X_NX_1)
=\varepsilon$. It is then easy to verify
\begin{eqnarray}
{\rm Tr}\prod_{j=1}^N\tilde{h}_{2j-1}
=
{\rm Tr}\prod_{j=1}^N\tilde{h}_{2j},
\end{eqnarray}
so that the enlarged representation necessarily admits an invertible KW duality. 
As shown in App. \ref{App:OddMajorana}, the trace difference is always an integer multiple of $2^N$, so that $\ell\in\mathbb Z_{\ge0}$. Thus the failure of invertibility in a given representation is not an intrinsic obstruction of the bond algebra itself: any sector imbalance can be cancelled by a finite enlargement of the representation. {This perspective is closely aligned with the recent construction \cite{PhysRevLett.134.130403}, where ancillary degrees of freedom provide a systematic mechanism for realizing duality operators as unitary linear-depth quantum circuits.} Here, for finite-dimensional representations of the Ising bond algebra, the obstruction to an invertible duality operator is captured by the trace mismatch $\Tr P_{\rm o}-\Tr P_{\rm e}$ and can always be removed by a finite extension of the representation.


\subsection{Ising chain and Majorana fermions}
We now turn to several familiar realizations of the Ising bond algebra, which serve to illustrate the unifying content of our result. The standard and twist-extended Ising chains, as well as the periodic and antiperiodic Majorana chains, emerge as distinct representations of the same algebra, with their respective duality properties captured by our criterion. For the $N$-site periodic quantum Ising chain, the bond operators are given by $h_{2j-1}=Z_j,\,h_{2j}=X_jX_{j+1}$, with the central elements $P_{\rm o}={\sf P}_z=Z_1\cdots Z_N$ and $P_{\rm e}=1$, which immediately rules out an invertible duality. The noninvertible duality operator in \eqref{eq:noninvKWD} assumes the well-known form 
\cite{seiberg2024majorana,Seiberg2024_NonInvertibleLSM}
\begin{eqnarray}\label{eq:noninvKWIsing}
    &&{\sf D}=e^{-\frac{2\pi{\rm i}N}{8}}\left(\prod_{j=1}^{N-1}e^{{\rm i}\frac{\pi}{4}Z_{j}}e^{{\rm i}\frac{\pi}{4}X_jX_{j+1}}\right)e^{{\rm i}\frac{\pi}{4}Z_{N}}\left(\frac{1+{\sf P}_z}{2}\right),\nonumber\\
    &&\qquad{\sf D}^\dagger {\sf D}=\left(\frac{1+{\sf P}_z}{2}\right),\quad{\sf D}^2=\left(\frac{1+{\sf P}_z}{2}\right)T,
\end{eqnarray}
where $T$ is the lattice translation operator, implementing $T{\cal O}_jT^{-1}={\cal O}_{j+1}$, with ${\cal O}_j=X_j,Z_j$. One now immediately finds that $\ell=1$ and $\varepsilon=-1$. Thus the trace obstruction is removed by adjoining a single antiperiodic Ising representation to the periodic one \cite{PRXQuantum.4.020357,PRXQuantum.5.010338,PhysRevB.108.214429,Seiberg2024_NonInvertibleLSM}. Equivalently, this direct sum can be realized locally by introducing an auxiliary spin and choosing $h_{2N}=X_N Z_0 X_1$. Indeed, in the $Z_0=\pm1$ sectors the boundary bond reduces respectively to $\pm X_NX_1$, so that the enlarged Hilbert space contains the periodic and antiperiodic Ising representations as two invariant sectors. Modifying the boundary term leads to $P_{\rm o}={\sf P}_z$ and $P_{\rm e}=Z_0$, which readily satisfy ${\rm Tr}\,P_{\rm o}={\rm Tr}\,P_{\rm e}=0$, thus ensuring an invertible duality. A possible choice for $\eta$ can be given as $\eta=(X_0+Y_0)X_1/\sqrt{2}$, which, in the basis of the auxiliary spin, leads to the duality operator 
\begin{eqnarray}
    {\cal D}=\begin{pmatrix}
        {\sf D} & X_1{\sf D}\\
        {\sf D}X_N & -{\rm i}X_1{\sf D}X_N
    \end{pmatrix},
\end{eqnarray}
Notably, we have exploited the $U(1)$-freedom of ${\cal D}_\pm$ in \eqref{eq:KWop} to arrive at this expression. One now has the algebra
\begin{eqnarray}\label{eq:extendedIsingKW}
    &&{\cal D}^\dagger{\cal D}=1,\quad{\cal D}^2={\cal T},\quad {\cal D}\hat{\cal P}=\hat{\cal P}{\cal D},\quad {\cal T}\hat{\cal P}=\hat{\cal P}{\cal T},\nonumber\\
    &&\text{with}\quad{\cal T}=\begin{pmatrix}
        T & 0 \\
        0 & Z_1T
    \end{pmatrix},\quad\hat{\cal P}=\begin{pmatrix}
        {\sf P}_z & 0 \\
        0 & -{\sf P}_z
    \end{pmatrix}.
\end{eqnarray}
The unitary duality thus squares to the corresponding lattice translation operators, with $T$ and $Z_1T$ implementing translations in the $N$-site periodic and antiperiodic Ising sectors, respectively. We note that a relation very similar to Eq. \eqref{eq:extendedIsingKW} was previously established in \cite{Seiberg2024_NonInvertibleLSM} using MPO construction. 

The local Hamiltonian densities of the Ising chain also admit a Majorana representation, with the spin-1/2 Pauli matrices represented in terms of Majorana fermions through the Jordan-Wigner transformation. In the fermionic language, the bonds become local Majorana bilinears. We now discuss the periodic Majorana chain, described by the densities $h_j=-{\rm i}\gamma_j\gamma_{j+1},~j=1,\ldots,2N$. It is easy to find that \eqref{eq:tracecond} is satisfied as ${\rm Tr}P_{\rm o}=0={\rm Tr}P_{\rm e}$, with $P_{\rm o}=-P_{\rm e}$, while $\eta$ may be chosen as $\gamma_1$. The resulting duality operator is given by ${\cal D}=\gamma_1\prod_{j=1}^{2N-1}(1+\gamma_j\gamma_{j+1})/\sqrt{2}$, which is precisely the periodic Majorana translation operator, satisfying ${\cal D}\gamma_j{\cal D}^{-1}=\gamma_{j+1}$. Since $\mathcal{D}$ exchanges the sectors $\mathcal{H}_{+-}$ and $\mathcal{H}_{-+}$, at the self-dual point $g=1$ the duality becomes a symmetry and enforces two-fold degeneracy of the spectrum \cite{PhysRevLett.117.166802}. Similarly, we can consider the antiperiodic Majorana chain, with the boundary term modified to $h_{2N}={\rm i}\gamma_{2N}\gamma_1$, while keeping the other densities as they are. This gives rise to the antiperiodic Majorana chain where one has $P_{\rm o}=P_{\rm e}$ and the invertible duality is found within the bond algebra itself. It remains to say that the so obtained duality is nothing but the antiperiodic translation. The KW duality of the periodic Ising chain can be given by projecting the antiperiodic and the periodic Majorana translation to the even $({\sf P}_z=1)$ and odd $({\sf P}_z=-1)$ parity sectors, respectively. However, only the projection to the even parity sector survives and gives rise to the noninvertible KW duality in \eqref{eq:noninvKWIsing}. 

\subsection{Order-disorder duality in a hidden Ising model.}
\label{sec:application}
These familiar realizations provide benchmarks for the general criterion. We now apply it to a nontrivial microscopic realization of the same bond algebra, where the resulting order-disorder duality is not apparent from the original spin variables. Consider the Hamiltonian  
\begin{eqnarray}\label{eq:Compass}
    &&H[J_z,J_x,\lambda_z,\lambda_x]=-\sum_{j=1}^N (J_zZ_{2j-1}Z_{2j}+J_xX_{2j}X_{2j+1})\nonumber\\
    &&\quad-\sum_{j=1}^N(\lambda_zZ_{2j}X_{2j+1}Z_{2j+2}+\lambda_x X_{2j-1}Z_{2j}X_{2j+1}),
\end{eqnarray}
with $X_{2N+1}\equiv X_1,Z_{2N+2}\equiv Z_2$ and $J_x,J_z,\lambda_x,\lambda_z\geq 0$. In the limit $\lambda_x,\lambda_z=0$, the Hamiltonian reduces to the well-known {\it quantum compass model} \cite{nussinov2015compass,brzezicki2007quantum,you2008quantum,shibata2019dissipative}, which  arises as effective description of systems with direction-dependent interactions. Finite $\lambda_x$ and $\lambda_z$ introduce direction-dependent three-spin cluster interactions of the $XZX$ and $ZXZ$ type. Similar three-spin terms have been explored previously in generalized compass models as a route to richer phase structure and low-energy behavior \cite{PhysRevB.93.214417,PhysRevB.95.224404}. The above Hamiltonian has an obvious strong-weak duality $H[J_z,J_x,\lambda_z,\lambda_x]\longleftrightarrow H[J_x,J_z,\lambda_x,\lambda_z]$, realized by a one-site translation followed by a global on-site Hadamard transformation. As a result, the Hamiltonian becomes self-dual at $J_x=J_z,\lambda_x=\lambda_z$. 
Beyond this manifest self-duality, we now show that when either $\lambda_x$ or $\lambda_z$ vanishes, the model additionally exhibits a hidden order-disorder duality. First consider $\lambda_z=0$, with a redefinition of the Hamiltonian densities
\begin{eqnarray}
    &h_{2j-1}=Z_{2j-1}Z_{2j},\quad h_{2j}=X_{2j-1}{\cal C}_jX_{2j+1},\nonumber\\
    &{\cal C}_j=\left(J_x X_{2j-1}X_{2j}+\lambda_x Z_{2j}\right)/(J_x^2+\lambda_x^2)^{1/2},
\end{eqnarray}
which leads to a realization of the bond algebra \eqref{eq:IsingBondAlgebra} \cite{sinha2026hidden}. The above Hamiltonian then can be rescaled by $(J_x^2+\lambda_x^2)^{1/2}$ to \eqref{eq:GenIsingBonHam} with $g={J_z}/{(J_x^2+\lambda_x^2)^{1/2}}$. The conserved parities $P_{\rm o}=\prod_{j=1}^{N}Z_{2j-1}Z_{2j},~ P_{\rm e}=\prod_{j=1}^{N}{\cal C}_j$ satisfy ${\rm Tr}\,P_{\rm o}={\rm Tr}\,P_{\rm e}=0$, thus ensuring an invertible duality. We can now construct the auxiliary operators $f_j$'s from \eqref{eq:auxferm}, identify a possible $\eta=Z_1X_2$, and use these to construct ${\cal D}$. Notably, the self-dual point is given by $g=1$ which in terms of the original couplings translates into $J_z=(J_x^2+\lambda_x^2)^{1/2}$.

Under the duality transformation $h_j\to h_{j+1}$, we have ${\cal D}(Z_{2j-1}\cdots Z_{2k}){\cal D}^{-1}=X_{2j-1}({\cal C}_j\cdots{\cal C}_k)X_{2k+1}$. It also maps the energy eigenstates at couplings $g$ and $1/g$ as $|E\rangle_{1/g}=
{\cal D}|E\rangle_{g}$. Furthermore, the local operators ${\cal C}_j$ satisfy $[{\cal C}_j,{\cal C}_k]=0$, ${\cal C}_j^2=1$, and commute with all the Hamiltonian densities $\{h_j\}$. Hence, they belong to the commutant of the bond algebra and the Hamiltonian can be diagonalized in a basis of simultaneous eigenstates of $\{{\cal C}_j\}$. However, for our choice of $\eta$, the operator ${\cal D}_-$ anticommutes with ${\cal C}_1$ and commutes with every other ${\cal C}_j$. Therefore, in the sectors labeled by the eigenvalues $\{{\cal C}_{j\neq 1}=\pm 1\}$, we have the order-disorder duality
\begin{eqnarray}
    \left\langle Z_{2j-1}Z_{2j}\cdots Z_{2k-1}Z_{2k} \right\rangle_g=\pm\left\langle X_{2j-1}X_{2k+1}\right\rangle_{1/g},
\end{eqnarray}
with $j\leq k$ and $j,k\neq 1$. It resembles the usual Ising order-disorder duality relation between the order $\langle X_jX_k\rangle$ and disorder $\langle Z_j\cdots Z_{k-1}\rangle$ correlations, with $j,k$ restricted to the odd sites and the sign $\pm$ depending on the particular sector under consideration. The complementary case $\lambda_x=0$ follows from the manifest translation-Hadamard duality and has hidden self-dual line $J_x=(J_z^2+\lambda_z^2)^{1/2}$.   

\section{Discussion.}
\label{sec:discussion}
We have shown that the finite-size realization of Kramers-Wannier duality is controlled by the representation of the underlying Ising bond algebra. For a general realization of this algebra, the canonical bond-algebraic construction yields a noninvertible duality on a closed finite chain. We have established that equality of the traces of the central elements $P_{\rm o,\rm e}$ is necessary and sufficient for extending the KW duality to an invertible unitary transformation on the full Hilbert space. Thus, invertibility is not an intrinsic property of the abstract bond-algebra automorphism alone, but depends on the representation in which it is realized. We emphasize that our criterion concerns the existence of an invertible unitary realization on a given finite-dimensional Hilbert space. This is complementary to recent constructions of ancillary-assisted unitary circuits \cite{PhysRevLett.134.130403} and to approaches realizing dualities as locality-preserving quantum cellular automata on the full operator algebra \cite{jones2026quantum}. Our trace criterion therefore characterizes unitary implementability at the level of finite-dimensional representations, without imposing circuit-depth or locality-preserving constraints on the duality operator. The resulting duality provides an exact correspondence between the spectra as $E(g)/g=E(1/g)$ and becomes a unitary symmetry at $g=1$. Since the duality exchanges the orthogonal sectors ${\cal H}_{+-}$ and ${\cal H}_{-+}$, at the self-dual point $g=1$, each energy with $\hat{\cal P}=-1$ becomes at least two-fold degenerate. This can be understood as the self-dual Hamiltonian $H(1)$ being supersymmetric when restricted to the odd parity sector. In fact, one can construct a supercharge \cite{PhysRevLett.117.166802} $Q=\sqrt{H_-/2}\,{\cal D}(1+(-1)^F)$, which squares to zero and anticommutes with its hermitian adjoint to yield the Hamiltonian $\{Q,Q^\dagger\}=2H_-$. Here $H_-=\hat{\cal P}_-H(1)\hat{\cal P}_-$ is the projection of the self-dual Hamiltonian $H(1)$ onto the parity-odd sector and shifted by a constant to ensure non-negativity of the energies. The operator $(-1)^F$ defines the $\mathbb{Z}_2$ grading in the parity-odd sector and may be identified with either $P_{\rm o}$ or $P_{\rm e}$, since the two are related by $\hat{\cal P}=P_{\rm o}P_{\rm e}=-1$ and are therefore not independent. Interestingly, the condition \eqref{eq:tracecond} can also be understood as the vanishing of a supersymmetric Witten index\cite{witten1982constraints}
\begin{eqnarray}
    {\cal I}_{\rm KW}(\beta)={\rm Tr}_-\left[(-1)^Fe^{-\beta H_-}\right],
\end{eqnarray}
where ${\rm Tr}_-[\,\cdot\,]={\rm Tr}\,[\hat{\cal P}_-\,\cdot\,]$ denotes trace over the parity-odd sector. For a unitary duality ${\cal D}$, satisfying ${\cal D}P_{\rm o,\rm e}{\cal D}^\dagger=P_{\rm e,\rm o}$ and $[{\cal D},H(1)]=0$, one obtains ${\cal I}_{\rm KW}(\beta)=0$. Subsequently, the condition \eqref{eq:tracecond} is recovered as ${\cal I}_{\rm KW}(0)=(1/2){\rm Tr}[P_{\rm o}-P_{\rm e}]=0$.

In recent years, Kramers–Wannier duality has been generalized to gauge theories, higher-form symmetries, and quantum many-body systems \cite{choi2022noninvertible,schafer2024ictp,li2026generalized}. Its connection with non-invertible symmetries and topological defects has provided new insights into critical phenomena and exotic phases of matter. It would therefore be interesting to investigate generalized KW dualities in this framework, as well as whether analogous criteria for invertibility persist for $p$-range bond algebras, in which each density anticommutes with its $p$ neighboring bonds.


\bibliography{refs}
\bibliographystyle{ieeetr}

\onecolumngrid
\appendix

\section{}
\label{App:OddMajorana}
We consider the representation of the complex Clifford algebra ${\rm CL}_{2N-1}(\mathbb{C})$, generated by $\{f_j\}$, satisfying
\begin{eqnarray}
    \{f_j,f_k\}=2\delta_{jk},\qquad j,k=1,\ldots,2N-1 .
\end{eqnarray}
It is well-known that the above algebra has two inequivalent irreducible representations of dimension $2^{N-1}$ \cite{das2014lie}. For an odd number of generators, the product of all the generators defines a central element of the Clifford algebra 
\begin{eqnarray}
    P_{\rm o}
    =
    (-{\rm i})^{N-1}f_1\cdots f_{2N-1},\qquad
    P_{\rm o}^2=1,
    \qquad
    [P_{\rm o},f_j]=0 .
\end{eqnarray}
Since $P_{\rm o}$ lies in the center of the algebra, Schur's lemma implies that in any irreducible representation it must act as a scalar multiple of the identity. Combining this with $P_{\rm o}^2=1$, we have the irreducible representations
\begin{eqnarray}
    \rho_\pm(P_{\rm o})=\pm\mathbb{1}_{2^{N-1}},\qquad \rho_\pm:{\rm CL}_{2N-1}(\mathbb{C})\to{\rm End}\left({\frak H}_\pm\right),
\end{eqnarray}
with ${\frak H}_\pm$ being the irreducible sectors with ${\rm dim}\,{\frak H}_\pm=2^{N-1}$. Let us consider a particular irreducible representation $\varrho(f_j)$ of the Clifford algebra with $\varrho(P_{\rm o})=\mathbb{1}_{2^{N-1}}$. Since every other irreducible representation $\rho_+(f_j)$ with $\rho_+(P_{\rm o})=\mathbb{1}_{2^{N-1}}$ is equivalent to $\varrho(f_j)$ and every $\rho_-(f_j)$ with $\rho_-(P_{\rm o})=-\mathbb{1}_{2^{N-1}}$ is equivalent to $-\varrho(f_j)$ \cite{das2014lie}, we can express all the irreducible representations as
\begin{eqnarray}
    \rho_+(f_j)=T\varrho(f_j)T^{-1},\qquad \rho_-(f_j)=-S\varrho(f_j)S^{-1},
\end{eqnarray}
with invertible operators $S,T$. For $f_j=f_j^\dagger$ describing Majorana fermions, we further require 
\begin{eqnarray}
    \rho_+(f_j)=\rho_+(f_j)^\dagger\implies \left[T^\dagger T,\varrho(f_j)\right]=0.
\end{eqnarray}
Since, $\varrho$ is an irreducible representation, by Schur's lemma we have $T^\dagger T\propto 1$. In a completely similar fashion, we can show $S^\dagger S\propto 1$. In other words, the operators $S,T$ can be rescaled to unitaries. 

Let us now consider a representation of the bond algebra on a Hilbert space ${\cal H}$ of dimension ${\rm dim}\,{\cal H}$. Therefore, the Clifford generators are represented by ${\rm dim}\,{\cal H}\times {\rm dim}\,{\cal H}$ dimensional matrices. The two central elements of the bond algebra $P_{\rm o,\rm e}$ provide a natural decomposition of the total Hilbert space as
\begin{eqnarray}
    {\cal H}={\cal H}_{++}\oplus{\cal H}_{--}\oplus{\cal H}_{+-}\oplus{\cal H}_{-+},\qquad {\cal H}_{\alpha\alpha'}=\Pi_{\alpha\alpha'}{\cal H},\qquad \Pi_{\alpha,\alpha'}=\frac{1+\alpha P_{\rm o}}{2}\frac{1+\alpha'P_{\rm e}}{2},\quad \alpha,\alpha'=\pm.
\end{eqnarray}
The Clifford generators $\{f_j\}$ commute with both $P_{\rm o}$ and $P_{\rm e}$. The unitary equivalence among the irreducible representations then allows us to choose a basis in which the representation of the fermions takes the form
\begin{eqnarray}\label{eq:repferm}
    \rho(f_j)=\begin{pmatrix}
        \mathbb{1}_{r_{++}}\otimes\varrho(f_j) & 0 & &\\
        0 & -\mathbb{1}_{r_{--}}\otimes\varrho(f_j) & &\\
        & & \mathbb{1}_{r_{+-}}\otimes\varrho(f_j) & 0\\
        & & 0 & -\mathbb{1}_{r_{-+}}\otimes\varrho(f_j)
    \end{pmatrix}.
\end{eqnarray}
Here $r_{\alpha\alpha'}$ is the multiplicity of irreducible representations in the invariant sector ${\cal H}_{\alpha\alpha'}$. The dimension of the each sector is then given by ${\rm dim}\,{\cal H}_{\alpha\alpha'}=r_{\alpha\alpha'}2^{N-1}$. The central elements subsequently can be written as
\begin{eqnarray}\label{eq:repcentral}
    \rho(P_{\rm o})=\begin{pmatrix}
        \mathbb{1}_{r_{++}}\otimes\mathbb{1}_{2^{N-1}} & 0 & &\\
        0 & -\mathbb{1}_{r_{--}}\otimes\mathbb{1}_{2^{N-1}} & &\\
        & & \mathbb{1}_{r_{+-}}\otimes\mathbb{1}_{2^{N-1}} & 0\\
        & & 0 & -\mathbb{1}_{r_{-+}}\otimes\mathbb{1}_{2^{N-1}}
    \end{pmatrix},\nonumber\\
    \rho(P_{\rm e})=\begin{pmatrix}
        \mathbb{1}_{r_{++}}\otimes\mathbb{1}_{2^{N-1}} & 0 & &\\
        0 & -\mathbb{1}_{r_{--}}\otimes\mathbb{1}_{2^{N-1}} & &\\
        & & -\mathbb{1}_{r_{+-}}\otimes\mathbb{1}_{2^{N-1}} & 0\\
        & & 0 & \mathbb{1}_{r_{-+}}\otimes\mathbb{1}_{2^{N-1}}
    \end{pmatrix}.
\end{eqnarray}
Now, imposing ${\rm Tr}\,\rho(P_{\rm o})-{\rm Tr}\,\rho(P_{\rm e})=0$ immediately gives 
\begin{eqnarray}
    {\rm Tr}\,\rho(P_{\rm o})-{\rm Tr}\,\rho(P_{\rm e})=(r_{+-}-r_{-+})2^N=0\quad\implies\quad r_{+-}=r_{-+}\equiv r.
\end{eqnarray}
This lets us introduce the operator
\begin{eqnarray}
    \eta_-=\begin{pmatrix}
        0 & 0 & &\\
        0 & 0 & &\\
        & & 0 & \mathbb{1}_{r}\otimes\mathbb{1}_{2^{N-1}}\\
        & & \mathbb{1}_{r}\otimes\mathbb{1}_{2^{N-1}} & 0
    \end{pmatrix},\qquad \eta_-^\dagger\eta_-=\eta_-\eta_-^\dagger=\begin{pmatrix}
        0 & 0 & &\\
        0 & 0 & &\\
        & & \mathbb{1}_{r}\otimes\mathbb{1}_{2^{N-1}} & 0\\
        & & 0 & \mathbb{1}_{r}\otimes\mathbb{1}_{2^{N-1}}
    \end{pmatrix}=\hat{\cal P}_-.
\end{eqnarray}
From \eqref{eq:repferm} and \eqref{eq:repcentral}, it is now easy to check that
\begin{eqnarray}
    \{\eta_-,\rho(f_j)\}=0,\qquad \{\eta_-,\rho(P_{\rm o})\}=0=\{\eta_-,\rho(P_{\rm e})\}.
\end{eqnarray}
Since such an $\eta_-$ exists for any representation with ${\rm Tr}\,\rho(P_{\rm o})={\rm Tr}\,\rho(P_{\rm e})$, we drop the label $\rho$ from now on. Interestingly, if we further have ${\rm Tr}\,P_{\rm o}=0={\rm Tr}\,P_{\rm e}$, one can show that $r_{++}=r_{--}=r'$, which leads to 
\begin{eqnarray}
    \eta=\begin{pmatrix}
        0 & \mathbb{1}_{r'}\otimes\mathbb{1}_{2^{N-1}} & &\\
        \mathbb{1}_{r'}\otimes\mathbb{1}_{2^{N-1}} & 0 & &\\
        & & 0 & \mathbb{1}_{r}\otimes\mathbb{1}_{2^{N-1}}\\
        & & \mathbb{1}_{r}\otimes\mathbb{1}_{2^{N-1}} & 0
    \end{pmatrix},\qquad \eta^\dagger\eta=\eta\eta^\dagger=1. 
\end{eqnarray}
As can be checked, it again satisfies 
\begin{eqnarray}
    \{\eta,\rho(f_j)\}=0,\qquad \{\eta,\rho(P_{\rm o})\}=0=\{\eta,\rho(P_{\rm e})\}.
\end{eqnarray}

\end{document}